\documentclass[letter]{aa} % for the letters
\usepackage{graphicx}
\usepackage{txfonts}
\begin{document}
   \title{Inhomogeneities on the surface of 21 Lutetia, the asteroid target of the Rosetta mission
\thanks{Based on observations made with the Italian Telescopio Nazionale 
Galileo (TNG) operated on the island of La Palma by the Fundacion Galileo 
Galilei of the INAF (Istituto Nazionale di Astrofisica) at the Spanish
Observatorio del Roque de los Muchachos of the Instituto de Astrofisica
de Canarias (DDT program partially performed during program AOT18/TAC21).}}

   \subtitle{Ground-based results before the Rosetta fly-by}

   \author{D. Perna \inst{1,2,3},
	  E. Dotto \inst{3},
          M. Lazzarin \inst{4},
          S. Magrin \inst{4},
          M. Fulchignoni \inst{2,5},
          M. A. Barucci \inst{2},
          S. Fornasier \inst{2,5},
          S. Marchi \inst{4},
          C. Barbieri \inst{4}
}

\offprints{D. Perna}

\institute{Dipartimento di Fisica, Universit\`a di Roma Tor Vergata, Via della Ricerca Scientifica 1, 00133 Roma, Italy\\
        \email{dperna@oa-roma.inaf.it}
\and LESIA, Observatoire de Paris, 5 Place J. Janssen, 92195 Meudon Principal Cedex, France
\and INAF -- Osservatorio Astronomico di Roma, Via Frascati 33, 00040 Monte Porzio Catone (Roma), Italy
\and Dipartimento di Astronomia, Universit\`a di Padova, Vicolo dell'Osservatorio 3, 35122 Padova, Italy
\and Universit\'{e} Paris Diderot -- Paris 7, 4 rue Elsa Morante, 75013 Paris, France
}

   \date{Received ...; accepted ...}

% \abstract{}{}{}{}{}
% 5 {} token are mandatory

  \abstract
   {
In July 2010 the ESA spacecraft Rosetta will fly by the main belt asteroid
21 Lutetia. Several observations of this asteroid have been performed so far, 
but its surface composition and nature are still a matter of debate.
For a long time Lutetia was supposed 
to have a metallic nature due to its high IRAS albedo. 
Later on it has been suggested that the asteroid has a
surface composition similar to primitive carbonaceous chondrite meteorites, 
while further observations proposed a possible genetic link with
more evolved enstatite chondrite meteorites.
}
   {We performed visible spectroscopic observations of 21 Lutetia in November 2008 at the Telescopio Nazionale Galileo (TNG, La Palma, Spain)
to make a decisive contribution to solving the conundrum of its nature. 
   }
   {Thirteen visible spectra were acquired at different rotational phases and subsequently analyzed. 
   }
   {We confirm a narrow spectral feature at about 0.47-0.48 
$\mu$m which was already found by Lazzarin et al. (2009) in the spectra of Lutetia. 
We also confirm an earlier find of Lazzarin et al. (2004), who detected a spectral feature at about 0.6 $\mu$m
in one of their Lutetia's spectra. More remarkable is the difference
of our spectra though, which exhibit different spectral slopes between 0.6 
and 0.75 $\mu$m and, in particular, we found that  
up to 20\% of the Lutetia surface could have flatter spectra. 
   }
   {We detected a variation of the spectral slopes at different 
rotational phases that could be interpreted as possibly due to 
differences in the chemical/mineralogical composition 
as well as to inhomogeneities of the structure of the Lutetia's surface
(e.g., to craters or albedo spots) in the southern hemisphere.}

\titlerunning {21 Lutetia: Ground-based results before the Rosetta fly-by}
\authorrunning{D. Perna et al.}

   \maketitle
%
%________________________________________________________________

\section{Introduction}

Rosetta is the ESA cornerstone mission devoted to the study of minor
bodies of the solar system. 
The main target is comet 67P/Churyumov-Gerasimenko that will be reached
in 2014, after three Earth and one Mars gravity assisted swing-bys.
During its journey the mission investigates also two main belt
asteroids, 2867 Steins (fly-by in September 2008) and 21 Lutetia
(fly-by in July 2010).

The asteroid 21 Lutetia was discovered in 1852 by H.~Goldschmidt at the Paris Observatory.
Zappal\`a et al. (1984) measured a rotational period of P=8.17$\pm$0.01 hours, a value later refined by
Torppa et al. (2003), who found P=8.165455~hours and
also computed the pole coordinates obtaining
a model with axis ratios a/b=1.2 and b/c=1.2.
Although several observations are available since
a couple of decades, the nature of this asteroid is still controversial. 
Radiometric measurements gave albedo values included in the range 
0.19--0.22 (IRAS, M\"{u}ller et al.
2006, Lamy et al. 2008).
More recently, Carvano et al. (2008) obtained a geometric albedo of 0.129, significantly
lower than all the previous estimations, and
explained the wide range of computed
albedo values as the evidence of  inhomogeneities on the
surface of Lutetia (e.g., one or
more large craters on the northern hemisphere).
Due to the first estimation of its albedo by IRAS,
Lutetia was supposed
to have a metallic nature (Barucci et al.
1987; Tholen 1989).
Later on it has been suggested that the asteroid has a
surface composition similar to primitive carbonaceous chondrite meteorites
(Howell et al. 1994; Burbine \& Binzel 2002; Lazzarin et al. 2004, 2009; 
Birlan et al. 2004; Barucci et al. 2005, 2008),
while Vernazza et al. (2009) proposed a possible genetic link with
more evolved enstatite chondrite meteorites.

On the basis of the observational evidence,
Lutetia appears to be an atypical puzzling asteroid, whose nature is still 
far from fully understood.
In order to enhance our knowledge of this unusual object, 
in November 2008 we performed visible spectroscopic observations of Lutetia
to investigate its surface and check if some
inhomogeneities are present.

\section{Observations and data reduction}

Observations were carried out at the 3.6 m
Telescopio Nazionale Galileo (TNG, La Palma, Spain). The observational circumstances are given in Table~1.

Visible spectroscopy was performed with the DOLORES
(Device Optimized for the LOw RESolution) instrument.
We used the low resolution blue (LR-B) grism, which cover the 0.34-0.81
$\mu$m range with a spectral dispersion of 2.5 \AA/px (http://www.tng.iac.es).
Spectra were taken
through a 2 arcsec wide slit oriented along the parallactic angle to
avoid flux loss due to the differential refraction.
Bias, flat--field, calibration lamp (He lines) and several spectra of solar 
analog stars (Landolt 98-978 and Hyades 64) were recorded during 
the observing run.
The spectra were reduced with the software packages Midas and IDL
using standard procedures (see, e.g., Dotto et al. 2009) which include
subtraction of the bias from the raw data, flat field correction, cosmic rays
removal, background subtraction, collapsing the two--dimensional spectra
into one dimension, wavelength calibration (using the He lamp's emission lines) and atmospheric extinction correction.
The reflectivity of Lutetia was then obtained by
dividing
its spectra by one of the solar analog star Hyades 64.
The resulting spectra were cut at 0.45 $\mu$m because at lower 
wavelengths the behavior is strongly affected by the spectra of the solar analogs. 
Table~1 also reports the coverage of the rotational phase of each of the
obtained spectra and the spectral slopes computed between 0.6 $\mu$m and 0.75 $\mu$m.
\begin{table}[t]
\caption{Observational circumstances: for each spectrum we report
the starting time of acquisition corrected for light-time, 
the exposure length, the airmass of Lutetia, the spectral slope 
computed between 0.6 $\mu$m and 0.75 $\mu$m, the coverage of the rotational phase, 
and the airmass of the used solar analog.}             % title of Table
\label{table:1}      % is used to refer this table in the text
\resizebox{\columnwidth}{!}{
\begin{tabular}{cccccccc}        % centered columns (4 columns)
\hline\hline                 % inserts double horizontal lines
\# & Date & UT$_{start}$     &  T$_{exp}$  & Airm. & Spec.            & Rot.  & Sol. an.\\
   &      &                  &             &       & slope            & phase & airm.\\
   &      &  (hh:mm)         &  (s)        &       & (\%/10$^3\AA$) &       &    \\
\hline                        % inserts single horizontal line
1 & 2008 Nov 27 &  23:39 &   8   &  1.07  & 1.5$\pm$0.5 & 0.018 & 1.04 \\
2 & 2008 Nov 27 &  23:53 &   30  &  1.05  & 1.0$\pm$0.5 & 0.046 & 1.04 \\
3 & 2008 Nov 28 &  00:49 &   40  &  1.01  & 1.4$\pm$0.5 & 0.161 & 1.04 \\
4 & 2008 Nov 28 &  01:33 &   40  &  1.02  & 1.8$\pm$0.5 & 0.251 & 1.04 \\
5 & 2008 Nov 28 &  02:23 &   40  &  1.07  & 2.2$\pm$0.5 & 0.353 & 1.04 \\
6 & 2008 Nov 28 &  03:05 &   40  &  1.14  & 2.0$\pm$0.5 & 0.438 & 1.04 \\
7 & 2008 Nov 28 &  03:54 &   40  &  1.30  & 2.1$\pm$0.5 & 0.538 & 1.04 \\
8 & 2008 Nov 28 &  04:32 &   40  &  1.50  & 1.7$\pm$0.5 & 0.616 & 1.04 \\
9 & 2008 Nov 29 &  22:18 &   40  &  1.23  & 0.7$\pm$0.5 & 0.731 & 1.02 \\
10 & 2008 Nov 29 &  23:12 &  40  &  1.09  & 1.4$\pm$0.5 & 0.841 & 1.02 \\
11 & 2008 Nov 29 &  23:13 &  30  &  1.09  & 1.6$\pm$0.5 & 0.843 & 1.02 \\
12 & 2008 Nov 30 &  00:07 &  30  &  1.03  & 1.7$\pm$0.5 & 0.953 & 1.02 \\
13 & 2008 Nov 30 &  00:57 &  40  &  1.01  & 1.7$\pm$0.5 & 0.056 & 1.02 \\
\hline
\end{tabular}}
\end{table}
   \begin{figure}[t]
   \centering
   \includegraphics[angle=0,width=11.1cm]{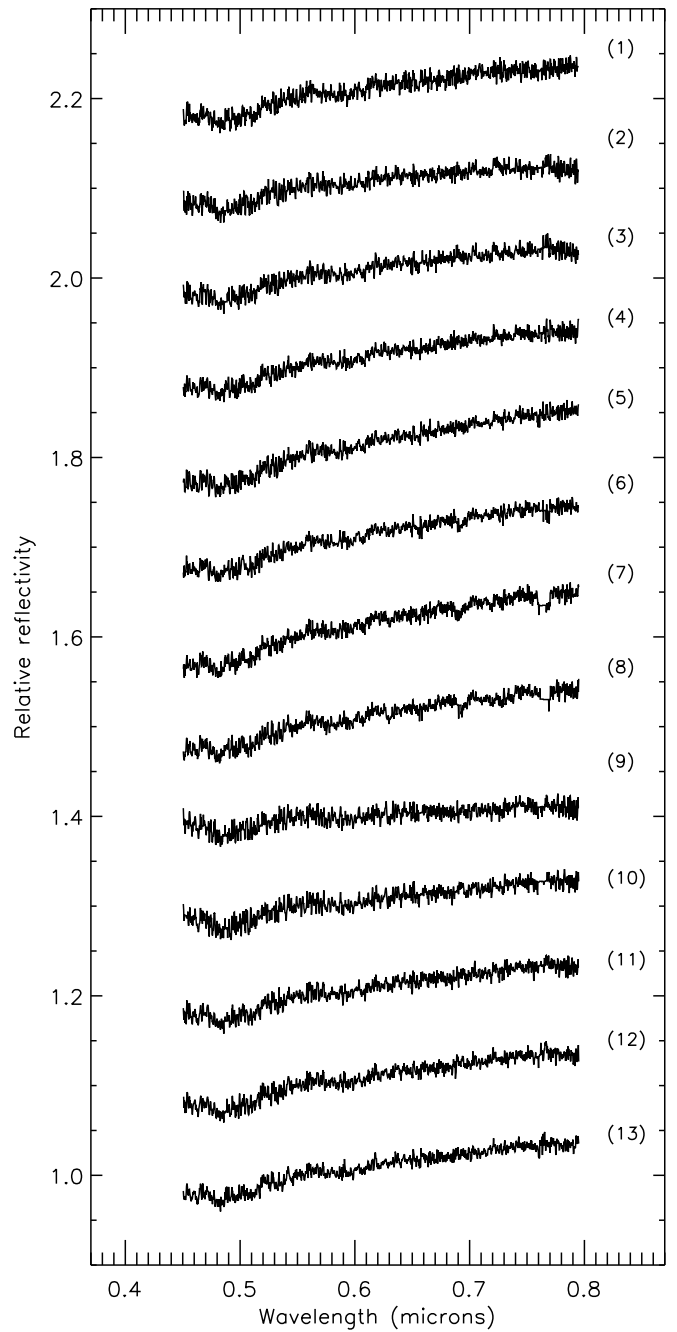}
   \caption{
Visible spectra of Lutetia normalized at 0.55 $\mu$m and shifted
by 0.1 in reflectivity for clarity. The numbers on the right correspond to the spectrum
number reported in Table~1 and Fig.~2.
               }
              \label{}%
    \end{figure}
   \begin{figure}[t]
   \centering
   \includegraphics[angle=0,width=11cm]{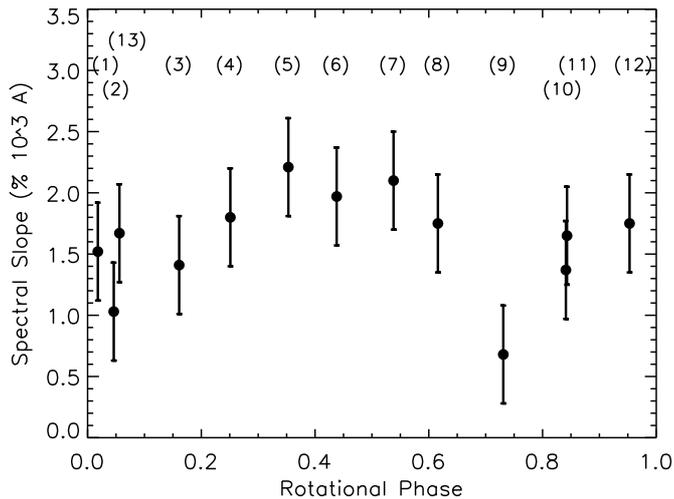}
   \caption{Spectral slope vs. rotational phase, computed considering a period of 8.165455 hours (Torppa et al. 2003).
               }
              \label{}%
    \end{figure}

\section{Obtained results and discussion} 

The obtained spectra are shown in Fig.~1.
The general behavior is similar in all of them, but some
differences are evident.

All spectra exhibit two weak
features at about 0.47--0.48 $\mu$m and around 0.6 $\mu$m.
Their depth relative to the continuum reflectance
is of about 1\% in all spectra, and does not exhibit any evident variation.
Bus and Binzel (2002a, b) observed features at about 0.48 $\mu$m and 0.6 $\mu$m
in the spectra of several main belt asteroids, which were therefore classified as
Xe types. Burbine et al. (2002) found that these spectral signatures are consistent with those detected in oldhamite, a mineral 
commonly found in the aubrites (enstatite achondrite meteorites), but which is also 
present in enstatite chondrites. 
Vernazza et al. (2009), who analyzed visible spectra and discussed 
the computed albedo values, suggested that the physical properties of Lutetia
are compatible with those of enstatite
chondrite meteorites.
Nevertheless, the shape of the feature at about 0.48 $\mu$m in the Lutetia spectra is
considerably different from that observable in Xe-type asteroids, probably indicating a dissimilar origin of the absorption.
Both features at $\sim$0.48 $\mu$m and 0.6 $\mu$m were already detected in the spectra of Lutetia (Lazzarin et al. 2004, 2009).
The signature at about 0.47--0.48 $\mu$m was detected 
by Lazzarin et al. (2009) and 
interpreted as due to spin-allowed
crystal field transitions in Ti$^{3+}$ in pyroxene M1 sites, although
Hazen et al. (1978)
suggested that in lunar pyroxenes this band
could be due to the superimposition
of Ti$^{3+}$and Fe$^{2+}$ effects.
The feature at about 0.6 $\mu$m was seen in one of the spectra published by
Lazzarin et al. (2004)
and it is generally attributed to charge transfer transitions in
minerals produced by aqueous alteration of anhydrous
silicates (Vilas et al. 1994), but can be also present in pyroxenes 
(see, e.g., Burns et al. 1976).
Additional constraints come from far-infrared Spitzer data of Lutetia, which do not confirm any similarity with 
enstatite chondrite meteorites (Barucci et al. 2008; Lazzarin et al. 2009), 
but suggest CO3-CV3 carbonaceous chondrites as the closest meteorite analogs.
Because the mineralogical composition of asteroids can be inferred only 
by taking into account multiwavelength observational data together with albedo 
values, the most plausible interpretation 
of the features detected in visible spectra of Lutetia at this point in time appears to be that
given by Lazzarin et al. (2004, 2009), while the analogy with enstatite chondrites seems unlikely.

Figure~2 plots the obtained spectral slopes (values reported in Table~1).
A variation through the rotational phase is evident and can be interpreted 
as due to inhomogeneities on the observed portion of the Lutetia's surface.
These inhomogeneities could be related to different chemical/mineralogical 
compositions, as well as to 
albedo spots or to craters exposing regions with different 
albedo/age.
In particular, a clear inhomogeneity emerges in correspondence
with spectrum \# 9, which could cover up to 20\% of the
surface we observed.

Considering the pole solution ($\lambda=39^\circ$, $\beta=3^\circ$) by 
Torppa et al. (2003), 
our spectra were obtained at an aspect angle $\xi \sim 30^\circ$.
Hence we can support the hypothesis
that one or more craters/inhomogeneities can be present in the
southern hemisphere of Lutetia.

\section{Conclusions}

An observational campaign of Lutetia was carried out on November 2008.
Visible spectroscopy was performed for
this main belt asteroid and 
thirteen spectra were acquired
at different rotational phases.
All of them exhibit absorption
features centered at about 0.47--0.48 $\mu$m and around 0.6
$\mu$m. 
The spectral slope shows a variation through the rotational phase,
suggesting that some inhomogeneities must be
present in a portion of up to 20\% of the 
observed surface of Lutetia in its southern hemisphere. 
%These inhomogeneities can be related to different chemical/mineralogical compositions, 
%as well as to the presence of {\bf albedo spots or of}
%craters {\bf exposing regions with different albedo/age}.
These differences in the acquired spectra  
can be due to inhomogeneities in the chemical/mineralogical composition, 
or to albedo spots or
craters exposing regions with different albedo/age.

These data are useful
in the assessment of the physical nature of this object, and will 
constitute a fundamental basis for the 
calibration, analysis and interpretation of the data acquired by the 
instruments onboard the Rosetta spacecraft.

\begin{acknowledgements}
We thank Emilio Molinari for the DDT shared between the groups of 
E. Dotto and M. Lazzarin. We are grateful to Francesca Ghinassi for 
help and useful discussions during the observations.
We also thank T.H. Burbine, referee of the manuscript, for providing
constructive comments and help in improving this paper.
\end{acknowledgements}
~\\[-1.1cm]

\end{document}